\documentclass[aps,showpacs,twocolumn]{revtex4}

\usepackage{epsfig}
\usepackage{epsf}
\usepackage{bm}                 % for bold math symbols
\usepackage{amsmath}
\usepackage{amssymb}
\usepackage{dcolumn}
\usepackage{graphicx}

\input amssym.def
\input amssym

\font\bbold=msbm10

\newcommand{\BR}{\mbox{\bbold R}}

\newcommand{\IN}{{\cal H}_{\rm in}}
\newcommand{\OUT}{{\cal H}_{\rm out}}
\begin{document}

\title{Signaling and the Black Hole Final State}

\author{Ulvi Yurtsever} \email{Ulvi.Yurtsever@jpl.nasa.gov}

\affiliation{Quantum Computing Technologies Group, Jet Propulsion 
Laboratory, California Institute of Technology \\ Mail Stop 126-347, 
4800 Oak Grove Drive, Pasadena, California 91109-8099}

\author{George Hockney} \email{George.Hockney@jpl.nasa.gov}

\affiliation{Quantum Computing Technologies Group, Jet Propulsion 
Laboratory, California Institute of Technology \\ Mail Stop 126-347, 
4800 Oak Grove Drive, Pasadena, California 91109-8099}

\date{\today}

\begin{abstract}
In an attempt to restore the unitarity of the evaporation process,
Horowitz and Maldacena~\cite{HM} recently proposed a
boundary-condition constraint for the final quantum state of an
evaporating black hole at its singularity. Gottesman and
Preskill~\cite{GP} have argued that the proposed constraint must lead to
nonlinear evolution of the initial (collapsing) quantum state. Here we
show that in fact this evolution allows signaling, making it
detectable outside the event horizon with entangled-probe experiments
of the kind we proposed recently~\cite{YH}. As a result the
Horowitz-Maldacena proposal may be subject to terrestrial tests.
\end{abstract}

\pacs{03.67.-a, 03.65.Ud, 04.70.Dy, 04.62.+v}

\maketitle 

We begin with a brief review of the final-state boundary condition for
evaporating black holes as proposed in~\cite{HM} and further elucidated
in~\cite{GP}. In the semiclassical approximation,
the overall Hilbert space for the evaporation process can
be treated as a decomposition
\begin{equation}
{\cal H} = {\cal H}_M \otimes {\cal H}_F = {\cal H}_M
\otimes {\cal H}_{\rm in} \otimes {\cal H}_{\rm out} \; ,
\end{equation}
where ${\cal H}_M$ denotes the Hilbert space of the quantum field that
constitutes the collapsing body, and ${\cal H}_F$ is the Hilbert space
in which the quantum-field fluctuations around the background spacetime
determined by the ${\cal H}_M$ quantum state live. The separation of
${\cal H}$ into ${\cal H}_M$ and ${\cal H}_F$ reflects the semiclassical
nature of the treatment in a fundamental way. Moreover, the
fluctuation Hilbert space ${\cal H}_F$ can be further decomposed as
${\cal H}_F = \IN \otimes \OUT$, where $\IN$ and $\OUT$
denote the Hilbert spaces of
fluctuation modes confined inside and outside the event horizon,
respectively. Before evaporation, the quantum state $|\; \rangle
\in {\cal H}$ of the complete system  can be written as a product
\begin{equation}
|\; \rangle = |\psi_0 \rangle_M \otimes |0_U\rangle \; ,
\end{equation}
where $|\psi_0 \rangle_M \in {\cal H}_M$
is the initial wave function of the collapsing matter,
and the Unruh vacuum $|0_U \rangle \in 
{\cal H}_F = \IN \otimes \OUT$ is the
maximally entangled state
\begin{equation}
|0_U\rangle = \frac{1}{\sqrt{N}}
\sum_{k=1}^{N} | k_{\rm in} \rangle \otimes |k_{\rm out} \rangle
\; .
\end{equation}
Here $N$ is the common dimension (the number of degrees of
freedom necessary to completely describe the internal state of the black hole)
of all three Hilbert spaces ${\cal
H}_M$, $\IN$, and $\OUT$, and $\{ |k_{\rm in}\rangle \}$ and
$\{ |k_{\rm out} \rangle \}$, $k=1,2,\cdots,N$,
are fixed orthonormal bases for $\IN$ and $\OUT$, respectively.
After the hole evaporates completely, the ``final" Hilbert space is simply
$\OUT$, and the usual semiclassical arguments inevitably imply a mixed
state $\rho_{\rm out}$ as the endpoint of complete evaporation (see
Fig.\,1 in~\cite{YH} and the associated discussion), revealing
that the transition $|\psi_0 \rangle_M \mapsto \rho_{\rm out}$ is
manifestly non-unitary.

The Horowitz-Maldacena proposal (HM) imposes a boundary condition on the
final quantum state at the black-hole singularity by demanding that it
be equal to
\begin{equation}
|\Phi\rangle
\equiv U^{\dagger} \left[ \frac{1}{\sqrt{N}} \sum_{j=1}^{N} | j_M \rangle
\otimes |j_{\rm in} \rangle \right] \in {\cal H}_M \otimes \IN
\end{equation}
where $\{ |j_M \rangle \}$ is an orthonormal basis for ${\cal H}_M$,
and $U: {\cal H}_M \otimes \IN \rightarrow {\cal H}_M \otimes \IN$ is a
unitary transformation. More precisely, HM states the following:

{\noindent \it There exists a unitary map
$U: {\cal H}_M \otimes \IN \rightarrow {\cal H}_M \otimes \IN$ such that
with $|\Phi\rangle \in {\cal H}$ defined as in Eq.\,(4),
the state $|\; \rangle$ [Eqs.\,(2)--(3)] evolves after
complete evaporation as
\begin{equation}
|\; \rangle \longmapsto \alpha \, P_{|\Phi\rangle \otimes \OUT}
\; | \; \rangle  \; ,
\end{equation}
where $\alpha \in \BR$ is a renormalization constant, and
$P_{|\Phi\rangle \otimes \OUT}$ denotes the projection onto the linear
subspace $|\Phi\rangle \otimes \OUT \equiv \{
|\Phi\rangle \otimes |v\rangle : \; |v\rangle \in \OUT \}$ of $\cal H$.}

The unitary operator $U$ describes the non-local evolution
of the black-hole quantum state near the singularity, as well as its
evolution in the semiclassical regime before the singularity; one would
expect a full quantum theory of gravity to be able
to completely specify this operator. To restore unitarity to the
transition map ${\cal H}_M \rightarrow \OUT$,
Horowitz and Maldacena~\cite{HM} further demand that $U$ be in
the form of a product corresponding to the absence
of entangling interactions between ${\cal H}_M$ and $\IN$:
\begin{equation}
U = S_1 \otimes S_2 \; ,
\end{equation}
where $S_1 : {\cal H}_M \rightarrow {\cal H}_M$ and $S_2 :
\IN \rightarrow \IN$ are unitary maps. To find the effective evolution
map ${\cal H}_M \rightarrow \OUT$ resulting from HM and the assumption
Eq.\,(6), start from the equality
\begin{equation}
\alpha \, P_{|\Phi\rangle \otimes \OUT} \;
(\, |\psi_0 \rangle_M \otimes |0_U\rangle \,)
= |\Phi\rangle \otimes |X_{\rm out} \rangle \; ,
\end{equation}
where $|X_{\rm out}\rangle$ is the state in $\OUT$ into which the initial
state $|\psi_0\rangle_M$ evolves after the evaporation. Contracting
both sides of Eq.\,(7) with $\langle \Phi |$ substituted from
Eq.\,(4)
\begin{eqnarray}
|X_{\rm out}\rangle = \frac{\alpha}{N}
\sum_{j=1}^{N}\langle j_M| \otimes \langle j_{\rm in}|
(S_1 \otimes S_2 ) \nonumber \\
|\psi_0\rangle_M \otimes \sum_{k=1}^{N} |k_{\rm in}\rangle \otimes
|k_{\rm out}\rangle \nonumber \\
 = \frac{\alpha}{N} \sum_{j=1}^{N}
\sum_{k=1}^{N} \langle j_M|S_1 |\psi_0\rangle_M
\, \langle j_{\rm in} | S_2 | k_{\rm in} \rangle
\; |k_{\rm out} \rangle  \; .
\end{eqnarray}
In terms of the basis components
$X_{{\rm out} \; j} \equiv \langle j_{\rm out}|X_{\rm out}\rangle$,
$\psi_{0 \; k} \equiv \langle k_M | \psi_0 \rangle_M$,
$S_{1 \; jk} \equiv \langle j_M | S_1 | k_M \rangle$, and
$S_{2 \; jk} \equiv \langle j_{\rm in} | S_2 | k_{\rm in} \rangle$,
Eq.\,(8) can be rewritten in the matrix form
\begin{equation}
X_{{\rm out }\; k} = \frac{\alpha}{N}
\sum_{l=1}^{N} ({S_2}^{T} S_1 )_{\, kl} \; \psi_{0 \; l} \; \; ,
\end{equation}
where $S^{T}$ denotes matrix transpose of $S$. Since the transpose of a
unitary matrix is still unitary, Eq.\,(9) shows that (i) the
renormalization constant $\alpha =N$, and (ii) the transformation
$|\psi_0\rangle_M \mapsto |X_{\rm out}\rangle$ is unitary.

However, as pointed out by Gottesman and Preskill~\cite{GP}, entangling
interactions between ${\cal H}_M$ and $\IN$ are unavoidable in any
reasonably generic gravitational collapse scenario. Consequently,
we cannot expect the unitary operator $U$ to
have the product form Eq.\,(6) in general. For a general unitary map
$U: {\cal H}_M \otimes \IN
\rightarrow {\cal H}_M \otimes \IN$, the vector $|\Phi\rangle$
defined by Eq.\,(4) is an arbitrary element in
${\cal H}_M \otimes \IN$, and Eq.\,(8) leads to
the more general linear expression
\begin{equation}
X_{{\rm out }\; k} = {\alpha}
\sum_{l=1}^{N} T_{\, kl} \; \psi_{0 \; l} \; \; 
\end{equation}
instead of Eq.\,(9). Here $T$ denotes the matrix
\begin{equation}
T_{kl} \equiv \frac{1}{\sqrt{N}} \langle \, \Phi \; |l_M \rangle
\! \otimes \! | k_{\rm in} \rangle 
\end{equation}
which is unconstrained except for $\sum_{kl} |T_{kl}|^2 =
1/N$. Only when $U$ has the product form Eq.\,(6) $T$ equals ($1/N$
times) a unitary matrix [Eq.\,(9)].
Note that the constant $\alpha$ is to be determined from the
condition that $|X_{\rm out}\rangle$ remains normalized.
After this renormalization,
we can express the transformation ${\cal H}_M \rightarrow \OUT$
described by Eq.\,(10) more succinctly in the form
\begin{equation}
X_{{\rm out }\; k} = \frac{1}{(\, \sum_{i} | \! \sum_{j}
T_{\, ij} \; \psi_{0\; j} |^2 \, )^{\frac{1}{2}}}
\sum_{l=1}^{N} T_{\, kl} \; \psi_{0 \; l} \; \; 
\end{equation}
where now $T$ is a {\it completely unconstrained},
arbitrary matrix~\cite{ftnote1}.
While it maps pure states to pure states,
the transformation ${\cal H}_M \rightarrow \OUT$ specified by Eq.\,(12)
is not only nonunitary, but it is in fact {\it nonlinear};
linearity is recovered (along with unitarity) only when $T$ is
proportional to a unitary matrix.

In a recent paper~\cite{YH}, we argued that nonlinear quantum evolution
inside an evaporating black hole might have observable consequences outside
the event horizon when an entangled system (whose coherence is carefully
monitored) partially falls into the hole. We also proposed a specific
experiment that should be able to detect the presence of such
signaling nonlinear maps via terrestrial quantum interferometry. We
will now show that the HM-class of nonlinear maps defined in
Eq.\,(12) in fact belong to this signaling class. Therefore, the HM
boundary-condition proposal can in principle be tested by terrestrial
experiments.

Let us assume a causal configuration as depicted in Fig.\,1
of~\cite{YH}, where a bipartite system $AB$ evolves to send its $B$-half
into a black-hole event horizon along a null geodesic, while the
$A$-half remains coherently monitored outside the horizon. We can
then further decompose the ``collapsing" Hilbert space ${\cal H}_M$ in
the form ${\cal H}_M = {\cal H}_A \otimes {\cal H}_B$, where ${\cal
H}_B$ now corresponds to all matter that falls into the black hole,
including the ``probe beam" $e$ of our trans-horizon Bell-correlation
experiment (cf.\ Fig.\,2
and the discussion following it in~\cite{YH}), and ${\cal H}_A$
corresponds to all matter that remains outside the horizon, including
the interferometer beams which are monitored in the laboratory. We
also identify the outgoing Hilbert space $\OUT$ with ${\cal H}_M$, which
amounts to specifying a unitary map $U_M : {\cal H}_M \rightarrow \OUT$
connecting orthonormal basis sets in the two spaces. With this
identification, the ``evaporation" map ${\cal H}_M \rightarrow \OUT$ can
be treated as a map sending ${\cal H}_M$ onto ${\cal H}_M$.
Reinterpreted thus,
the action of a general quantum map in the class defined by Eq.\,(12)
can be written as
\begin{equation}
\rho_{AB} \longmapsto
\frac{T \, \rho_{AB} \, T^{\dagger}}
{{\rm Tr} (T \, \rho_{AB} \, T^{\dagger})} \; 
\end{equation}
on any state $\rho_{AB}$ in ${\cal H}_M
={\cal H}_A \otimes {\cal H}_B$,
where $T: {\cal H}_A \otimes {\cal H}_B
\rightarrow {\cal H}_A \otimes {\cal H}_B$ is a
(nonsingular) general linear
transformation~\cite{ftnote1}. To satisfy the locality condition as
formulated in Eq.\,(14) of~\cite{YH}, the map $T$ must have the
product form
\begin{equation}
T = T_A \otimes T_B \; ,
\end{equation}
where $T_A : {\cal H}_A \rightarrow {\cal H}_A$
and $T_B : {\cal H}_B \rightarrow {\cal H}_B$ are general linear maps.
Since subsystem $A$ remains outside the event horizon, the evolution map
$T_A$ must remain unitary, and we can assume
(for simplicity and without loss of
generality) that $T_A = \mathbb{I}_A$. Then the quantum evolution
map Eq.\,(13) acting
on the Hilbert space ${\cal H}_M = {\cal H}_A \otimes {\cal H}_B$
takes the more transparent form
\begin{equation}
{\cal E}_{AB} \; : \; \rho_{AB} \longmapsto
\frac{{\pmb 1}_A \otimes {\cal T}_B \; ( \rho_{AB})}
{{\rm Tr} [{\pmb 1}_A \otimes {\cal T}_B \; ( \rho_{AB})]} \; ,
\end{equation}
where ${\pmb 1}_A = {\cal E}_A$ denotes the identity
map on states of ${\cal H}_A$, and
${\cal T}_B$ denotes the linear transformation (not a quantum map)
\begin{equation}
{\cal T}_B \; : \; \rho_{B} \longmapsto
T_B \, \rho_{B} \, {T_B}^{\dagger} \; 
\end{equation}
on states of ${\cal H}_B$. When $\rho_{AB}$ is a product state
$\rho_{AB}=\rho_A \otimes \rho_B$, the action of
${\cal E}_{AB}$ has the manifestly local form
\begin{equation}
{\cal E}_{AB}(\rho_{AB})= {\cal E}_A (\rho_A )
\otimes {\cal E}_B (\rho_B ) \; ,
\end{equation}
where ${\cal E}_A = {\pmb 1}_A$, and ${\cal E}_B$ is the nonlinear
quantum map
\begin{equation}
{\cal E}_{B} \; : \; \rho_{B} \longmapsto
\frac{{\cal T}_B (\rho_B)}{{\rm Tr}_B [{\cal T}_B (\rho_B )]}
=
\frac{T_B   \rho_{B} {T_B}^{\dagger}}
{{\rm Tr}_B ( T_B   \rho_{B} {T_B}^{\dagger})} \; 
\end{equation}
mapping ${\cal H}_B$-states onto ${\cal H}_B$-states (compare
Eq.\,(17) above with Eq.\,(14) of~\cite{YH}). By contrast,
when $\rho_{AB}$ is
entangled the action of ${\cal E}_{AB}$ does not
have the simple product form of Eq.\,(17).

The criterion for a quantum map ${\cal E}_{AB}$ to be signaling is
identified in~\cite{YH} (see Eq.\,(15) of~\cite{YH}
and the associated discussion there) as simply the condition that
\begin{equation}
{\rm Tr}_B \, [ \, {\cal E}_{AB} (\rho_{AB}) \, ]
\neq {\cal E}_A \, [ \, {\rm Tr}_B ( \rho_{AB}  ) \, ] \; 
\end{equation}
for {\it some} (necessarily entangled) state $\rho_{AB}$. Now consider
a class of entangled states
$\rho_{AB}$ in the form of a convex linear combination
\begin{equation}
\rho_{AB} = \lambda_1 \, \rho_A \otimes \rho_B + \lambda_2
\, \sigma_A \otimes \sigma_B \; ,
\end{equation}
where $\rho_A , \; \sigma_A$ , and $\rho_B , \; \sigma_B$ are
(normalized) states in
${\cal H}_A$ and ${\cal H}_B$, respectively, and $\lambda_1 >0,
\; \lambda_2 > 0 , \; \lambda_1 + \lambda_2 = 1$ are real coefficients.
Introduce the real numbers
\begin{eqnarray}
n_1 & \equiv & {\rm Tr}_B [{\cal T}_B (\rho_B )] = {\rm Tr}_B (T_B \rho_B
{T_B}^{\dagger}) \; , \nonumber \\
n_2 & \equiv & {\rm Tr}_B [{\cal T}_B (\sigma_B )] = {\rm Tr}_B (T_B \sigma_B
{T_B}^{\dagger}) \; .
\end{eqnarray}
The right-hand-side of Eq.\,(19) is simply ${\rm Tr}_B (\rho_{AB})$
(recall that ${\cal E}_A = {\pmb 1}_A$):
\begin{equation}
{\cal E}_A \, [ \, {\rm Tr}_B ( \rho_{AB}  )]
= \lambda_1 \rho_A + \lambda_2 \sigma_A \; ,
\end{equation}
while the left-hand-side is
\begin{equation}
{\rm Tr}_B \, [ \, {\cal E}_{AB} (\rho_{AB}) \, ]
= \frac{\lambda_1 \, n_1 \, \rho_A \; + \; \lambda_2 \, n_2 \, \sigma_A}
{\lambda_1 \, n_1 + \lambda_2 \, n_2} \; .
\end{equation}
But
\begin{equation}
\lambda_1 \rho_A + \lambda_2 \sigma_A \; \neq \;
\frac{\lambda_1 \, n_1 \, \rho_A \; + \; \lambda_2 \, n_2 \, \sigma_A}
{\lambda_1 \, n_1 + \lambda_2 \, n_2}
\end{equation}
{\it unless} at least one of the conditions: (i) $n_1 = n_2$, or
(ii) $\rho_A=\sigma_A$ holds. The condition (i) does not hold in general
unless the linear operator $T_B$ is unitary (or a scalar multiple of a
unitary operator), and condition (ii) does not hold in general unless
$\rho_{AB}$ is a product state. Therefore the nonlinear quantum map ${\cal
E}_{AB}$ defined by Eqs.\,(15)--(16) is in general in the signaling
class.

In summary, we have shown that the quantum
maps which likely characterize quantum evolution through
evaporating black holes according to the Horowitz-Maldacena~\cite{HM}
boundary-condition proposal are a signaling class.
It is clear that the HM-class of maps are
detectable with the same kind of apparatus we
described previously, namely the Zou-Wang-Mandel (ZWM)
interferometer depicted in Fig.\,2 of~\cite{YH}
[see Eqs.\,(11)--(13) of~\cite{YH} for a specific
example of the detection signal likely to arise from a HM-class
nonlinear map, in this case a $45^{\circ}$ shift
in the detector's interference fringes].
On the other hand, the precise nature of the signal
produced in the ZWM interferometer when the probe beam is sent
into an evaporating hole will depend on the nature of the unitary
operator $U$ characterizing the HM boundary condition, Eq.\,(4).
If, as predicted~\cite{HM,GP},
the operator $U$ involves nonlocal phases which oscillate chaotically
at Planckian frequencies near the singularity, then each ZWM
photon entering the hole is likely to experience a different nonlinear
evolution map ${\cal E}_{AB}$, and the observed signal will be an
average over such maps. It appears plausible
that this averaging will affect the local interference pattern back in
the laboratory by erasing relative phases and thus diminishing
fringe visibility. A detailed
discussion of this and other experimental questions will be found in a
forthcoming paper~\cite{moretocome}.

~~~

~~~

The research described in this paper was carried out
at the Jet Propulsion Laboratory under a contract with the National
Aeronautics and Space Administration (NASA), and
was supported by grants from NASA and the Defense
Advanced Research Projects Agency.

%\vspace{-0.2in}


\begin{thebibliography}{5}
\bibitem{ftnote1} The only constraint on the linear operator $T$
 entering the construction Eq.\,(12) or Eq.\,(13) is that
 $T$ must not have zero as an eigenvalue (e.g., projection-like
 operators are excluded), since with a singular
 $T$ this construction fails to give a well-defined quantum map.
\end{thebibliography}
\end{document}